\documentclass[prb,twocolumn,amsmath,amssymb,superscriptaddress]{revtex4-1}
\usepackage{epsfig, graphicx,graphics,amsmath,amssymb,float}
\usepackage[T1]{fontenc}
\usepackage[latin9]{inputenc}
\usepackage{amsmath}
\usepackage{amssymb}
\usepackage{xcolor}
\usepackage{amscd}
\usepackage{bm}
\usepackage{psfrag}

\usepackage{bbm} 

\usepackage[caption=false]{subfig}
\usepackage{subfig}
\usepackage{color}
\usepackage[bookmarks=true,colorlinks,linkcolor=blue,urlcolor=blue,citecolor=blue]{hyperref}

\newcommand{\be}{\begin{equation}}
\newcommand{\ee}{\end{equation}}
\newcommand{\bea}{\begin{eqnarray}}
\newcommand{\eea}{\end{eqnarray}}

\newcommand{\mc}{\mathcal}

\begin{document}

\title{Bound states in two-dimensional Fermi systems  with quadratic band touching }
%\title{Exchange-driven bound states in Fermi systems with quadratic band touching}
%\title{Bound states in two-dimensional Fermi systems with quadratic band touching and dipolar interactions}
%\title{Bound states in two-dimensional Fermi systems with quadratic band touching and spin-exchange interactions}
%\title{Formation of Bound States in Topologically Protected Quadratic Bands Touching Systems}% Force line breaks with \\

\author{Fl\'avio L. N. Santos}
\affiliation{Departamento de F\'isica, Universidade Federal de Minas Gerais, C. P. 702,  Belo Horizonte, MG, 30123-970, Brazil}
\affiliation{Universit\'e Paris-Saclay, CNRS, Laboratoire de Physiques des Solides, 91405, Orsay, France}
\author{M\^onica A. Caracanhas}
\affiliation{Instituto de F\'isica de S\~ao Carlos, Universidade de S\~ao Paulo, S\~ao Carlos, SP, 13560-970, Brazil}
\author{M. C. O. Aguiar}
\affiliation{Departamento de F\'isica, Universidade Federal de Minas Gerais, C. P. 702, Belo Horizonte, MG, 30123-970,  Brazil}
\affiliation{Universit\'e Paris-Saclay, CNRS, Laboratoire de Physiques des Solides, 91405, Orsay, France}
\author{Rodrigo G. Pereira}
\affiliation{International Institute of Physics and Departamento de F\'isica Te\'orica
e Experimental,  Universidade Federal do Rio Grande do Norte, 
Natal, RN, 59078-970, Brazil}

\date{\today}% It is always \today, today,

\begin{abstract}
The formation of bound states between mobile impurity particles and fermionic atoms  has been demonstrated in spin-polarized Fermi gases with attractive interspecies interaction.  We investigate bound states of  mobile impurities    immersed in a two-dimensional   system with a symmetry-protected quadratic band touching. In addition to the standard $s$-wave  interaction, we consider an anisotropic dipolar exchange  interaction that locally breaks point group symmetries. Using  a weak-coupling renormalization group approach and a ladder approximation for the impurity-fermion propagator,  we establish that the number of bound states can be controlled by varying the anisotropy of the exchange interaction. Our results show that the degeneracy and momentum dependence of the binding energies reflect some distinctive properties of the  quadratic band touching. 

\end{abstract}

\maketitle

%\tableofcontents

\section{\label{intro}Introduction}

Topological semimetals with  quadratic band touching (QBT) in two dimensions constitute    examples of gapless band structures protected by point group and time reversal symmetries \cite{FradkinBook,Sun}. Microscopic  models exhibiting  QBT  have been proposed and studied on the checkerboard and kagome lattices \cite{Sun,Liu2010,Uebelacker2011,Dora2014}. Unlike Dirac points in   graphene, two-dimensional QBT points have a nonvanishing  density of states  and their  effective action  is scale invariant with dynamical exponent $z=2$ \cite{FradkinBook}. This makes the QBT unstable  against   weak short-range  interactions and leads to phase transitions where at least one symmetry is spontaneously broken. As a consequence, anomalous quantum Hall and nematic semimetal phases  were predicted based on a perturbative renormalization group (RG) approach and mean-field theory \cite{Sun}, and  were recently   investigated in  numerical studies \cite{Sur2018,Zeng2018}.  Experimental realizations of QBT systems  in optical lattices have also been discussed \cite{Olschlager2012,Sun2012,Li2016}.  

In this work we consider a (pseudo-)spin-$1/2$ fermionic model where a single spin-down fermion  interacts with   a  QBT system  of majority, spin-up fermions. This limit of extreme population imbalance  has received considerable attention in the context of cold atomic realizations of  Fermi polarons \cite{Massignan2014,Zollner2011,Mathy2011,Parish2011,Klawunn2011,Schmidt2012}, where mobile impurity atoms are dressed by   particle-hole excitations of the Fermi gas in which they are immersed.  The quasiparticle properties of Fermi polarons have been measured using   radio-frequency spectroscopy  \cite{SchirotzekPRL2009,KoschorreckNature2012,KohstallNature2012,Scazza2017}. Beyond the conventional polaron picture,   mobile impurities can   probe exotic properties of many-body systems such as  topological phase transitions  \cite{Grusdt2016,Camacho2019,Qin2019,Grusdt2019} and quasiparticle breakdown associated with  quantum criticality  \cite{Caracanhas2013,Punk2013,Pereira,yan2019bose}. 

In Ref. \cite{Pereira},   the fate of a polaron in a QBT system was shown to depend on the particle-hole asymmetry of the band structure. If the effective mass of the upper band (above the QBT point) is larger than that of the lower band, a  repulsive $s$-wave impurity-fermion interaction decreases logarithmically with decreasing energy scale, giving rise to a marginal Fermi polaron. On the other hand, if the lower band has   larger effective mass, the effective interaction increases at low energies, driving  the quasiparticle weight to zero and bringing about  an emergent orthogonality catastrophe   \cite{Pereira}.

The purpose of this paper is twofold: First, we generalize the  model  of Ref. \cite{Pereira} to include a long-range spin exchange interaction between the mobile impurity and the majority fermions. The motivation comes from dipolar quantum gases \cite{Lahaye2009}, in which   spin exchange has been demonstrated experimentally \cite{Yan2013,Hazzard2014}. In these systems, the spatial anisotropy of the dipolar interaction can be controlled by varying the direction of the molecular electric  dipole  moments.  We show that in the low-energy limit the anisotropic spin exchange generates an impurity-fermion interaction that locally breaks point group symmetries. This modifies the renormalization group flow of the effective couplings in the quantum impurity model. We find a regime in which a bare repulsive interaction    becomes effectively attractive at low energies. Second, we study the formation of bound states in analogy with  the corresponding phenomenon  in  two-dimensional Fermi gases with attractive   interactions \cite{Zollner2011,Mathy2011,Parish2011,Klawunn2011,Schmidt2012}.   We find that   the spectrum of an impurity coupled to a QBT system can exhibit zero, one or two bound states depending on the  relative strength of the $s$-wave contact interaction and the symmetry-breaking interaction due to   anisotropic   exchange. In particular, for an attractive  $s$-wave  interaction and no anisotropic   exchange, there are two bound states which become degenerate   for vanishing total momentum. Turning on a small anisotropic interaction, the degeneracy point can move to  finite momenta along specific directions determined by the QBT Hamiltonian.

The remainder of the paper  is organized as follows: In Sec. \ref{model}, we present the microscopic model on the checkerboard lattice and the effective field theory in the continuum limit. In Sec. \ref{sec:RG}, we analyze the interacting model  using a  perturbative RG approach, which reveals the existence of a crossover regime where the effective coupling  changes sign. In Sec. \ref{sec:spectral}, we calculate the two-particle propagator and the associated pair spectral function in the ladder approximation, and  discuss  the different regimes for the formation of bound states. Our concluding remarks can be found in Sec. \ref{sec:conclusion}. The Appendix  contains   expressions for functions that appear in the RG equations  and some discussion about the two-body problem with one particle near the QBT.

\section{Model\label{model}}
\begin{figure}[t]
{\includegraphics[width=0.7\columnwidth]{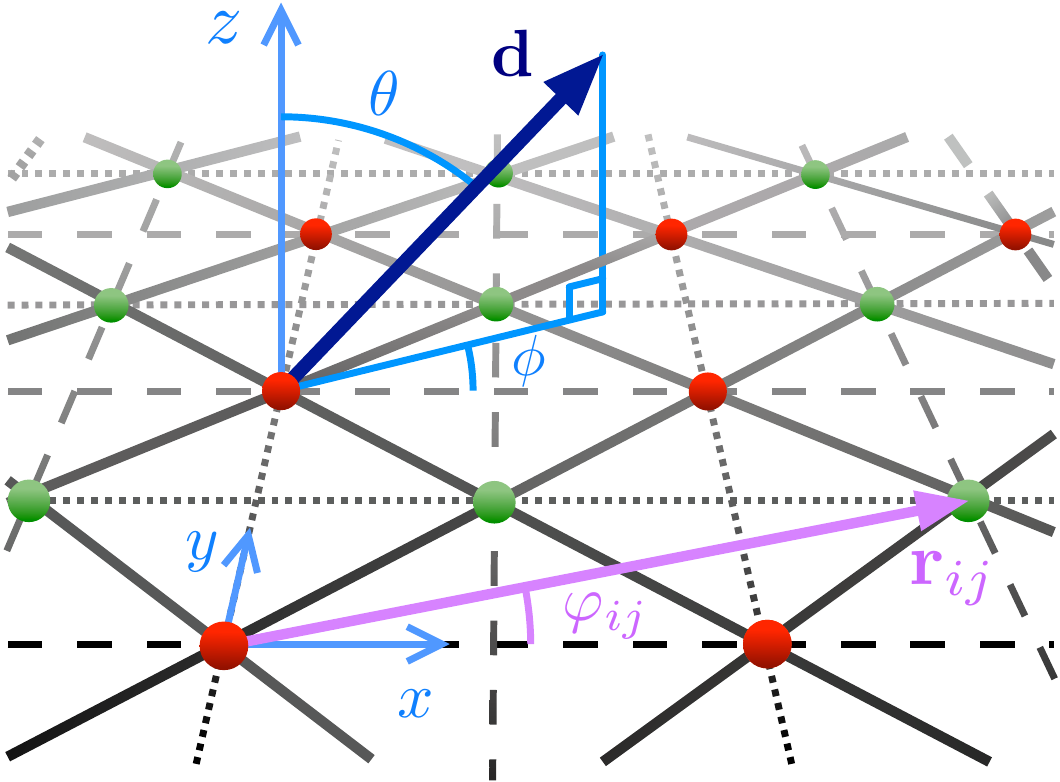}}
\caption{Checkerboard lattice.  Solid lines represent the nearest-neighbor hopping $t$ between sites in sublattice A (red) and B (green). Dashed and dotted lines represent next-nearest-neighbor hopping $t'$ and $t''$, respectively. The spin exchange interaction depends on the direction of   the dipolar moment $\mathbf d$, parametrized by the polar angle $\theta$  (with respect to the $z$ axis, perpendicular to the lattice plane) and the azimuthal angle $\phi$ (measured from the $x$ axis). A vector $\mathbf r_{ij}$ connecting two lattice sites forms an angle $\varphi_{ij}$ with the $x$ axis.}\label{rededispersao}
\end{figure}

We start with the model
\bea
H&=&-\sum_{\langle ij\rangle}t_{ij}(c^\dagger_{i\uparrow}c^{\phantom\dagger}_{j\uparrow}+c^\dagger_{i\downarrow}c^{\phantom\dagger}_{j\downarrow})+U\sum_{i}n_{i\uparrow}n_{i\downarrow}\nonumber\\
&&+\frac{J_\perp}4\sum_{  i\neq j}V_{ij}(S_i^+S_j^-+S_i^-S_j^+).\label{Hlattice}
\eea
Here, $c^\dagger_{j\alpha}$ creates a fermion  at site $j$ in one of two internal states, labeled by $\alpha=\uparrow,\downarrow$, and $n_{j\alpha}=c^\dagger_{j\alpha}c^{\phantom\dagger}_{j\alpha}$.  The hopping parameters $t_{ij}$ are defined on the checkerboard lattice. While the  nearest-neighbor hopping $t$ is uniform,   the next-nearest-neighbor hopping is either $t'$ or $t''$ depending on the sublattice and the direction of the link, as illustrated  in Fig. \ref{rededispersao}. For two next-nearest-neighbor sites in the A (B) sublattice, the hopping parameter is $t'$ along the $x$  ($y$) direction, but $t''$ along the $y$ ($x$) direction. In addition to the on-site Hubbard repulsion $U>0$, we consider a dipolar exchange interaction    \cite{Gorshkov2011,Zou2017} written in terms of   spin  operators $S_j^+=c^\dagger_{j\uparrow}c^{\phantom\dagger}_{j\downarrow}$ and $S_j^-=c^\dagger_{j\downarrow}c^{\phantom\dagger}_{j\uparrow}$.  The geometrical factor\be
V_{ij}= \frac{1-3(\hat{\mathbf d}\cdot \hat {\mathbf r}_{ij})^2}{|\mathbf r_{ij}|^3}
\ee
depends on the relative position  $\mathbf r_{ij}=\mathbf r_i-\mathbf r_j$ between sites. Here $\hat{\mathbf d}$ is a unit vector parallel to   the quantization axis,   set by  the direction of     the  polarized dipole moments \cite{Gorshkov2011}.  This type of exchange interaction  was  realized   using two rotational states of  polar molecules in optical lattices \cite{Yan2013}.  In terms of the angles shown in  Fig. \ref{rededispersao}, we can write $\hat{\mathbf d}\cdot \hat {\mathbf r}_{ij}=\sin\theta\cos(\phi-\varphi_{ij})$, where $\theta$ and $\phi$ are the angles of the $\mathbf d$ vector and $\varphi_{ij}$ is the angle between $\mathbf r_{ij}$ and the $x$ axis. Note that for  $\theta\neq0,\pi$ the  strength of the dipolar exchange interaction depends on the direction of $\mathbf r_{ij}$.

\begin{figure}[t]
{\includegraphics[width=0.7\columnwidth]{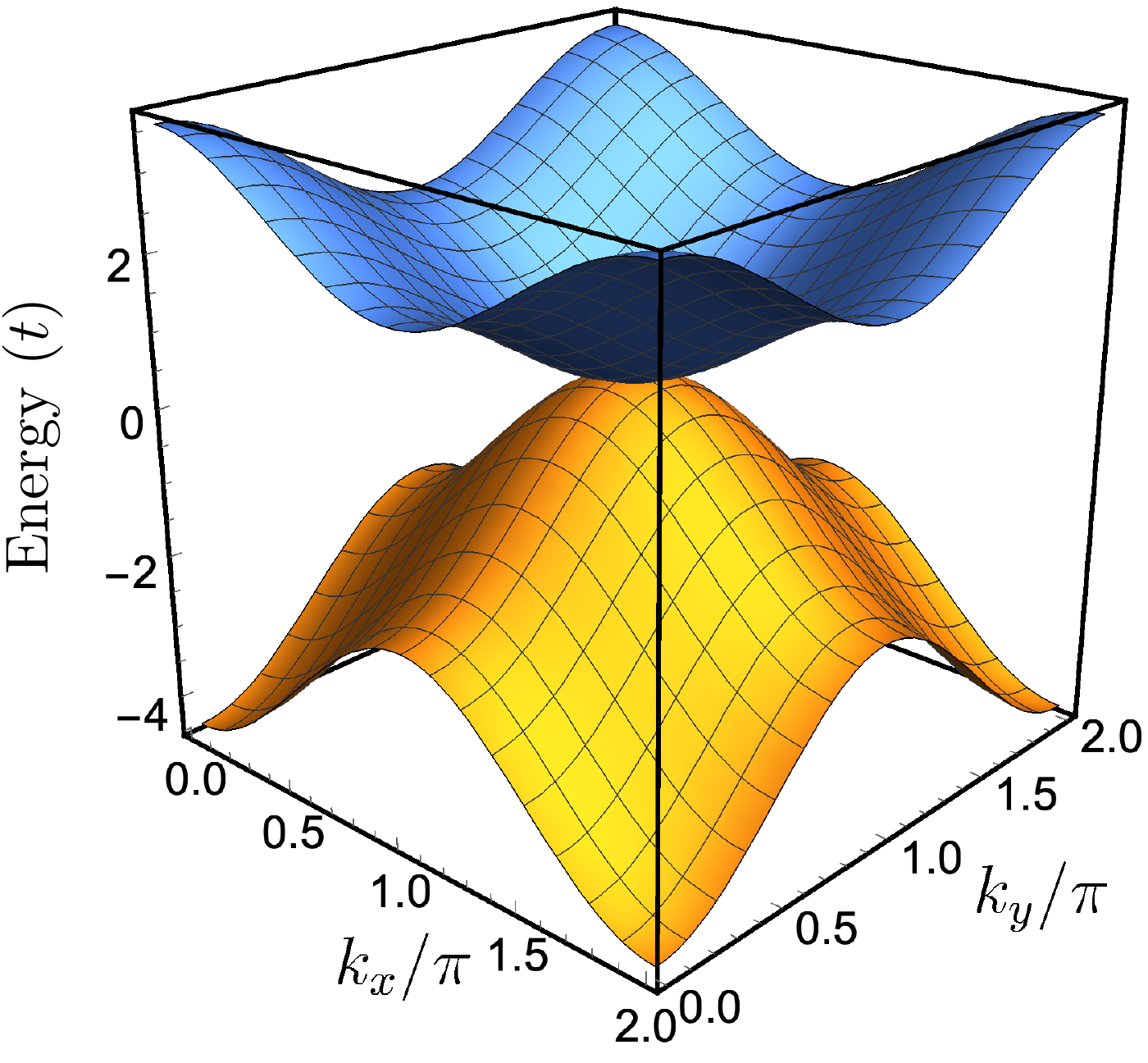}}
\caption{Band structure for the noninteracting checkerboard lattice model showing the quadratic band touching at $\mathbf Q=(\pi,\pi)$. Here we set $t'=0.6t$ and $t''=-0.4t$.}\label{bandstructure}
\end{figure}

In the noninteracting case, $U=J_\perp=0$, we can diagonalize the Hamiltonian using the mode expansion \bea
c_{j\alpha}&=&\left\{\begin{array}{ll}\frac1{\sqrt {N_s}}\sum_{\mathbf k}a_{\mathbf k\alpha}e^{i\mathbf k\cdot \mathbf R_j},&\quad j\in\textrm{A}\\
\frac1{\sqrt {N_s}}\sum_{\mathbf k}b_{\mathbf k\alpha}e^{i\mathbf k\cdot (\mathbf R_j+\boldsymbol\delta)},&\quad j\in\textrm{B}\end{array}\right.,
\eea
where $\mathbf R_j$ are positions on the square lattice with  lattice spacing set equal to $1$,  $N_s$ is the number of unit cells of the checkerboard lattice, and $\boldsymbol \delta=(\hat{\mathbf x}+\hat{\mathbf y})/2$ connects two sites in the same unit cell. The noninteracting Hamiltonian has the form
$H_0=\sum_{\mathbf k,\alpha}c^\dagger_{\mathbf k\alpha}\mc{H}_0(\mathbf k)c^{\phantom\dagger}_{\mathbf k\alpha}$, with \bea
\mc H_0(\mathbf k)&=& -(t'+t'')(\cos k_x+\cos k_y)\openone \nonumber\\ 
&&-(t'-t'')(\cos k_x-\cos k_y)\sigma^z\nonumber\\
&&-4t\cos(k_x/2)\cos(k_y/2)\sigma^x.
\eea
Here $c_{\mathbf k\alpha}=(a_{\mathbf k \alpha},b_{\mathbf k \alpha})$ is a two-component spinor and $\sigma^{x},\sigma^{y},\sigma^{z}$ are Pauli matrices acting in the sublattice space. The noninteracting Hamiltonian has a    C$_4$ rotational symmetry  corresponding to $\sigma^y\mc H_0(k_x,k_y)\sigma^y=\mc H_0(k_y,2\pi-k_x)$. In addition, $H_0$ is invariant under complex conjugation, equivalent to time reversal  in  sectors of the Fock space with fixed   $N_\uparrow=\sum_{j }c^\dagger_{j\uparrow}c^{\phantom\dagger}_{j\uparrow}$ and $N_\downarrow=\sum_j c^\dagger_{j\downarrow}c^{\phantom\dagger}_{j\downarrow}$. For $|t'+t''|<|t|$ and $|t'+t''|<|t'-t''|$, the band structure has a  QBT point at the corner of the Brillouin zone, $\mathbf Q=(\pi,\pi)$ \cite{Sun}, as illustrated  in Fig. \ref{bandstructure}. This  QBT point does not require fine tuning, since it carries  Berry phase $\pm 2\pi $   and is protected by C$_4$ and time reversal symmetries.  

Let us focus on the single-impurity model with $N_{\uparrow}=N_s$ and $N_\downarrow=1$ in the thermodynamic limit $N_s\to\infty$. In this case, the Fermi level of the spin-up (majority) fermions lies at the QBT point. We can describe their low-energy excitations  by expanding around momentum $\mathbf Q$. Hereafter we assume   $t''<0$ and $t=t'-t''>0$, in which case the dispersion around the QBT point becomes isotropic in the continuum limit \cite{Sun,Pereira}. By contrast, the low-energy limit for the impurity is obtained by expanding around the bottom of the lower band, at $\mathbf k=0$. The non-interacting Hamiltonian in the continuum limit becomes, up to a constant, \be
H_0=\int d^2r \left[\Psi^\dagger(\mathbf r) h_0(\mathbf r)\Psi(\mathbf r)-d^\dagger(\mathbf r)\frac{\nabla^2}{2M}d(\mathbf r)\right],\label{H0cont}
\ee 
where $\Psi(\mathbf r)=(\psi_A(\mathbf r),\psi_B(\mathbf r))^t$ is the two-component spinor associated with  the majority fermions and $d(\mathbf r)$ is the mobile impurity field with effective mass $M=(2t')^{-1}$. The operator \bea
h_0(\mathbf r)&=&\frac{m_+-m_-}{4m_+m_-}\openone \nabla^2+\nonumber\\ 
&&+\frac{m_++m_-}{4m_+m_-}\left[\sigma^z(\partial_x^2-\partial_y^2)+2\sigma^x\partial_x\partial_y\right]
\eea  
involves the effective masses  in the vicinity of the QBT point:  $m_+=[2(t-t')]^{-1}$ and   $m_-=(2t')^{-1}$ for the upper and lower bands, respectively.

We now switch on the interactions in the weak coupling regime $U,|J_\perp|\ll t$. The interacting Hamiltonian in the continuum limit has the form $H=H_0+H_{int}$, with $H_0$ given in Eq. (\ref{H0cont}) and the impurity-fermion interaction given by\be
H_{int}=\frac{4\pi}{m_+}\int d^2r \, \Psi^\dagger(\mathbf r)(g\openone +g_\perp\sigma^x) \Psi(\mathbf r)d^\dagger(\mathbf r)d(\mathbf r),\label{Hint}
\ee
where we define the dimensionless couplings 
\bea
g&=&\frac{m_+}{8\pi}\left[U-\kappa J_\perp \left(\frac{3}2\sin^2\theta-1\right)\right],\nonumber\\
g_\perp&=&-\frac{3m_+}{4\pi}\kappa_\perp J_\perp \sin^2\theta\sin(2\phi).\label{gamma1}
\eea
The latter stem from the Fourier transform of the on-site   and dipolar   exchange interactions and  contain the constants
\bea
\kappa&=&\frac32\zeta(3)-2\sum_{m=1}^\infty\sum_{n=1}^\infty\frac{(-1)^{m+n}}{(m^2+n^2)^{3/2}}\approx 1.322, \\
\kappa_\perp&=&\sum_{m=0}^{\infty}\sum_{n=0}^{\infty}\frac{(-1)^{m+n}\left(m+\frac12\right)\left(n+\frac12\right)}{\left[\left(m+\frac12\right)^2+\left(n+\frac12\right)^2\right]^{5/2}}\approx 1.312,\nonumber
\eea
where $\zeta(s)$ is the Riemann zeta function. 

We interpret   $g$  in Eq. (\ref{Hint}) as the usual $s$-wave scattering amplitude between the impurity and the majority fermions, whereas the new interaction $g_\perp$ scatters fermions between different sublattice states. 
Note that $g_\perp$ depends on the spatial anisotropy of the exchange interaction, and it vanishes when  the dipolar moment is polarized along the $z$ axis. 
In fact, the    $g_\perp$ interaction  breaks the C$_4$ symmetry, which in the continuum limit  becomes $\Psi(x,y)\mapsto \sigma^y\Psi(y,-x)$.   Importantly, both $g$ and $g_\perp$ are local interactions at  the position of the   mobile impurity and there are no interactions between majority fermions in the bulk. Thus, the single-impurity model allows us to explore the effects of a local symmetry-breaking interaction without destabilizing the QBT.

\section{Renormalization group analysis\label{sec:RG}} 

Short-range interactions are known to be marginal perturbations of two-dimensional semimetals with a QBT \cite{Sun,Murray2014,Pereira}. To treat the interactions  within perturbation theory,  we introduce  the   impurity Green's function     \be
G_d(\mathbf r,\tau)=-\langle T_\tau d(\mathbf r,\tau)d^\dagger(\mathbf 0,0)\rangle,
\ee
where  $d(\mathbf r,\tau)=e^{H\tau}d(\mathbf r)e^{-H\tau}$ is the impurity field  evolved in imaginary time, $T_\tau$ denotes time ordering with respect to $\tau$,  and the expectation value is calculated  in the ground state   with $N_\downarrow=0$.  To zeroth order in the interactions, we have the noninteracting Green's function in momentum-frequency domain:
\be
G^{(0)}_d(\mathbf k,i\nu)=\frac1{i\nu-k^2/(2M)}.
\ee
For the majority  fermions, we define  the  matrix Green's function 
\be
\mathbbm G=\left(\begin{array}{cc}
\mc G_{AA}&\mc G_{AB}\\
\mc G_{BA}&\mc G_{BB}
\end{array}\right),
\ee 
with components
\be
\mc G_{ll'}(\mathbf r,\tau)=- \langle T_\tau \psi^{\phantom\dagger}_l(\mathbf r,\tau) \psi^{\dagger}_{l'}(\mathbf 0,0)\rangle,
\ee
where    $l=A, B$ is the sublattice index.  
The Fourier-transformed noninteracting Green's function reads 
\bea
\mc G^{(0)}_{ll'}(\mathbf p,i\nu)&=&\left\{[i\omega\openone-\mc H_0(\mathbf Q+\mathbf p)]^{-1}\right\}_{ll'}\nonumber\\
&=&\sum_{\lambda=\pm}\frac{U_{l\lambda}(\mathbf p)U_{l'\lambda}(\mathbf p)}{i\nu-\lambda p^2/(2m_\lambda)}.
\eea
Here  $U_{l\lambda}(\mathbf p)$, with $\lambda=\pm$ the band index,  are the matrix elements  of the unitary  transformation that diagonalizes $h_0(\mathbf p)=\mc H_0(\mathbf Q+\mathbf p)$  with $|\mathbf p|\ll1$.  Due to the Berry phase associated with the QBT, $U(\mathbf p)$ depends on the angle $\varphi_{\mathbf p}=\arctan(p_y/p_x)$, in the form  \be
U(\mathbf p)=U(\varphi_{\mathbf p})=\left(\begin{array}{cc}
\sin\varphi_{\mathbf p}&\cos\varphi_{\mathbf p}\\
-\cos\varphi_{\mathbf p}&\sin\varphi_{\mathbf p}
\end{array}\right).\label{Umatrix}
\ee

\begin{figure}[t]
{\includegraphics[width=0.85\columnwidth]{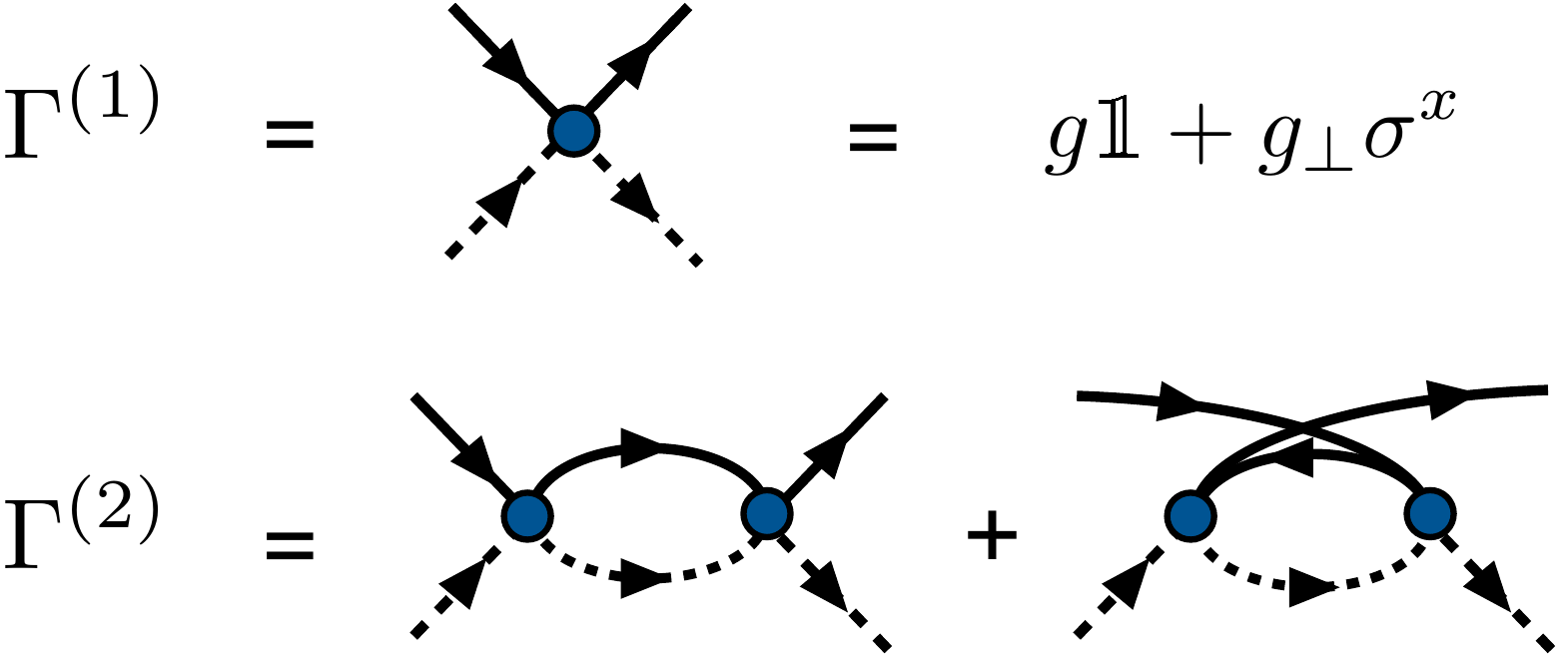}}
\caption{Effective interaction vertex at tree level ($\Gamma^{(1)}$) and at one-loop level ($\Gamma^{(2)}$). Solid lines represent the bare propagator for majority fermions, while dashed lines represent the  impurity propagator. The matrices in the interaction vertex act on the fermion sublattice degree of freedom.}\label{vertex}
\end{figure}

We analyze the effects of the impurity-fermion interaction using a weak-coupling Wilsonian RG approach \cite{Shankar,Cardy1996}. We derive the RG equations for the coupling constants at one-loop level and for the impurity effective mass and quasiparticle weight at two-loop level by integrating out high-energy fermion states in the  momentum shell   $\Lambda(1-d\ell)<p^2/(2m_+)<\Lambda$, where $\Lambda$ is the ultraviolet  cutoff and $d\ell \ll 1$ is the infinitesimal parameter in the RG step. For instance, the diagrams that contribute to the renormalization of the interaction vertex   are shown in Fig. \ref{vertex}.  We obtain the set of coupled RG equations:
 \begin{eqnarray}
   \frac{dg}{d\ell}&=& \frac{\left(\mu_--\mu_+\right)Z_d}{m_+}(g^2+g_{\perp}^2),\nonumber\ \\
  \frac{dg_{\perp}}{d\ell}&=& \frac{2(\mu_--\mu_+)Z_d }{m_+}gg_{\perp},\label{RGeqns}\\
   \frac{dZ_d}{d\ell}&=&-\frac{2\mu_-\mu_+Z_d}{m_+}\left[ 
   g^2 F_1(r_+,r_-)+g_{\perp}^2 F_2(r_+,r_-)\right],\nonumber\\
   \frac{dM}{d\ell}&=&\frac{2(\mu_-\mu_+)^{3/2}}{m_+}\left[ 
   g^2 F_3(r_+,r_-)+g_{ \perp}^2 F_4(r_+,r_-)\right],\nonumber 
 \end{eqnarray}
where $Z_d$ is the impurity quasiparticle weight, $\mu_\pm=m_\pm M/(M+m_\pm)$ are reduced masses, and $r_\pm =m_\pm/M$ are mass ratios.
The functions $F_i(r_+,r_-)$, with $i=1,\dots,4$,  are given  in terms of  integrals in Appendix \ref{AppFunctions} and return positive values of order 1. Note that  bulk properties, such as the effective   masses $m_+$ and $m_-$ for the majority fermions, are not renormalized in the single-impurity problem. 

The case $g>0$ and $g_\perp=0$ was studied in Ref. \cite{Pereira}. In this case, $g$ can be marginally relevant or irrelevant   depending on the difference between the effective masses $m_+$ and $m_-$. The reason is that  the two one-loop diagrams in the vertex renormalization (see  Fig. \ref{vertex}) have opposite signs.  For $m_->m_+$,   the diagram with a   hole propagator in the loop dominates and  the repulsive impurity-fermion interaction   flows to strong coupling. Ultimately, the quasiparticle weight $Z_d$ vanishes and  the effective impurity mass $M$ diverges logarithmically in the low-energy limit \cite{Pereira}.

\begin{figure}[t]
\centering
\includegraphics[width=0.9\columnwidth]{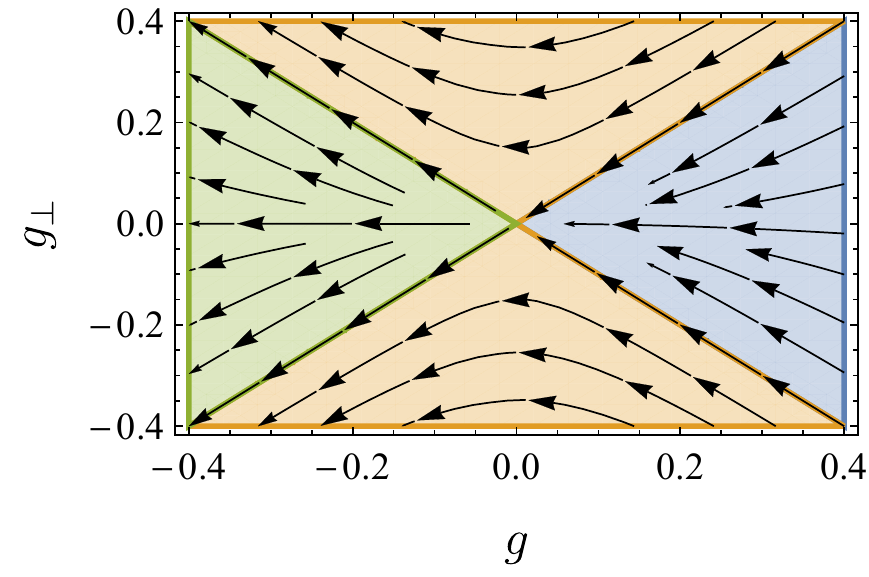}
\caption{Renormalization group  flow of the couplings in the single-impurity model with $m_-<m_+$. In the crossover region $|g|<|g_\perp|$ (orange), an initially repulsive $s$-wave scattering amplitude $g>0$ can change sign and become   attractive.}\label{rgflow1}
\end{figure}

Here we are interested in the case $m_-<m_+$, in which the diagram with a fermionic particle propagator in the loop dominates the vertex renormalization. The RG flow diagram for the couplings  $g$ and $g_\perp$ in Fig. \ref{rgflow1} reveals three regions with qualitatively different behavior. For $|g_\perp|<g$ (blue region in Fig. \ref{rgflow1}), the interaction is marginally irrelevant. As a result, in the low-energy limit the impurity decouples from the fermionic bath and one recovers Fermi polaron behavior with logarithmic corrections \cite{Pereira}. When we start off with an attractive interaction in the regime   $g<-|g_\perp|$ (green region in Fig. \ref{rgflow1}), the system exhibits monotonic flow to strong coupling.  Finally and most remarkably, for $|g|<|g_\perp|$ (orange region in Fig. \ref{rgflow1}), we observe a crossover from weak repulsive interaction  to strong attractive interaction, $g<0$. Our goal in the following will be to analyze the fate of the impurity in the latter two regions.

\section{Pair spectral function\label{sec:spectral}}

The flow of the effective couplings to strong attraction signals the formation of bound states between the impurity and a majority fermion. In two dimensions, at least one bound state exists in the two-body problem for an arbitrarily weak attractive interaction \cite{Mathy2011,Parish2011,Klawunn2011,Schmidt2012}.   To investigate the presence of  bound states, we consider  the pair creation  operator\be
P^\dagger(\mathbf r_j)=c^\dagger_{j\uparrow}c^\dagger_{j\downarrow}.
\ee
We then define the two-particle propagator as a matrix in sublattice space, with components\bea
\Pi_{ll'}(\mathbf R,\tau)&=&-(-1)^{s_l+s_{l'}}2\langle T_\tau P(\mathbf R+s_l\boldsymbol\delta,\tau)
P^\dagger(s_{l'} \boldsymbol\delta,0)\rangle,\nonumber\\ \label{2pprop}
\eea
%\bea \Pi_{ll'}(\mathbf R,\tau)&=&-(-1)^{s_l+s_{l'}}2\langle T_\tau P(\mathbf R+s_l\boldsymbol\delta,\tau) \nonumber\\ & &\times  P^\dagger(s_{l'} \boldsymbol\delta,0)\rangle,\label{2pprop} \eea
%
where   $\mathbf R$ is a position vector in sublattice A and   $s_A=0$, $s_B=1$. At low energies, we can work with   the two-particle propagator in the continuum limit:
\be
\Pi_{ll'}(\mathbf r,\tau)=-\langle T_\tau \psi_l^{\phantom\dagger}(\mathbf r,\tau)d(\mathbf r,\tau)d^\dagger(\mathbf 0,0)\psi_{l'}^\dagger(\mathbf 0,0)\rangle,
\ee 
where the factor of $(-1)^{s_l+s_{l'}}2$ in Eq. (\ref{2pprop}) gets cancelled in the projection of $c_{j\downarrow}$ onto the impurity field. Taking  the Fourier transform, \be
\Pi_{ll'}(\mathbf q,i\omega)=\int  d^2rd\tau\, e^{i\omega\tau}e^{-i\mathbf q\cdot \mathbf r}\, \Pi_{ll'}(\mathbf r,\tau),
\ee
and the analytic continuation $i\omega\to \omega+i0^+$, we define the pair  spectral function \be
\mc A_{\textrm{pair}}(\mathbf q,\omega)=-2\textrm{Im}\{\textrm{Tr}[\Pi(\mathbf q,\omega+i0^+)]\}.
\ee  
When interpreting the result for $\mc A_{\textrm{pair}}(\mathbf q,\omega)$ in the continuum limit in terms of the original lattice model, we must recall that zero energy corresponds to the impurity at the bottom of the lower band and the spin-up fermion at the QBT point. 

 \begin{figure}[t]
\centering
\includegraphics[width=.95\columnwidth]{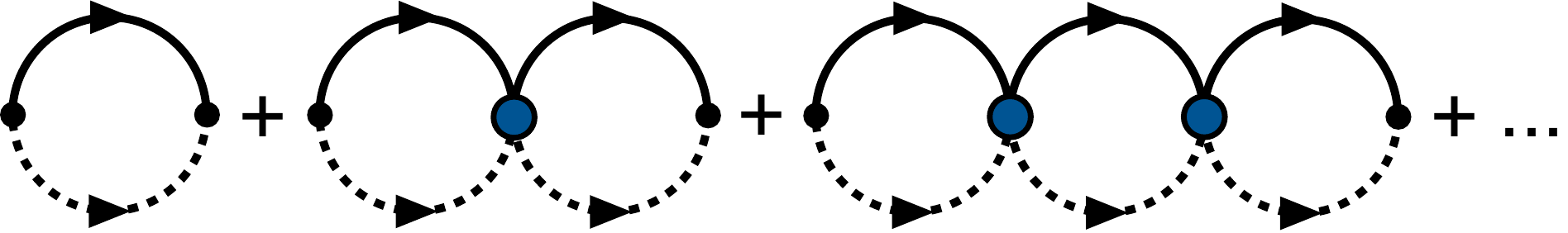}
\caption{Feynman diagrams included in the ladder approximation for the two-particle propagator.  The convention for the interaction vextex  and for impurity and fermion propagators is the same as in Fig. \ref{vertex}. }\label{ladder}
\end{figure}

We calculate the two-particle propagator in the ladder   approximation \cite{Mahan,Massignan2014}. This approximation is justified because, according to the RG analysis in Sec. \ref{sec:RG},  for  $m_-<m_+$ the perturbative expansion is dominated by  diagrams with a particle propagator in the   loops. The ladder series is illustrated in Fig. \ref{ladder}. %To zeroth-order in the interaction, we have 
The diagrams involve the bare two-particle propagator\bea
\Pi_0(\mathbf{q},i\omega)&=& \int \frac{d^2pd\nu}{(2\pi)^3} \mathbb G^{(0)}(\mathbf p+\mathbf q,i\omega+i\nu) G^{(0)}_{d}(-\mathbf p,-i\nu)\nonumber\\
&=&\frac{\mu_+}{4\pi}\left\{ \log\left(\frac{W-i\omega}{\Omega(q)-i\omega}\right)\openone\right.\nonumber \\ 
  & &-\left[ 1+ \frac{Mq^2-2iM^2\omega}{\mu_+q^2}\log\left(\frac{i\omega-\Omega(q)}{  i\omega-\frac{q^2}{2M} }\right) \right] \nonumber\\
 & & \times \left[\cos{(2\varphi_{\mathbf q})}\sigma^z+\sin{(2\varphi_{\mathbf q})}\sigma^x  \right]\Big\},\label{Iq}
\eea
where $W$ is a high-energy cutoff and 
 $\Omega(q) = \frac{q^2}{2(M+m_+)}$ is the lower threshold of the two-particle  continuum in the absence of interactions, corresponding to the minimum energy for   one fermion and the impurity carrying total momentum $\mathbf q$.  Note that $\Pi_0(\mathbf{q},i\omega)$ contains ``$d$-wave'' terms with nontrivial dependence on the angle $\varphi_{\mathbf q}$.

The two-particle propagator  is determined by   the Bethe-Salpeter equation in the ladder approximation 
\be
\Pi(\mathbf q,i\omega)=\Pi_0(\mathbf q,i\omega)\left[\openone +(g\openone +g_\perp\sigma^x)\Pi(\mathbf q,i\omega)\right],
\ee
which we solve by summing up a   geometric series of matrices. 
 We can identify   bound states by searching for  poles of $\Pi(\mathbf{q},\omega)$  below the two-particle continuum. We find two possible bound state dispersion relations, $E^{\pm}_{bs}(\mathbf q)$, given by the solutions to    \bea
  E^{\pm}_{bs}&=&\frac{\Omega(q)}{1-e^{X_\pm(\mathbf q,E^{\pm}_{bs})}} +\frac{W}{1- e^{-X_\pm(\mathbf q,E_{bs}^\pm)}}, \label{Enbs}
  \eea
  where \bea 
 X_{\pm}(\mathbf{q},E_{bs}^\pm)&=&\frac{(1+r_+)g}{g^2-g_\perp^2}  \pm \frac{1+r_+}{g^2-g_\perp^2}    \nonumber\\
 &&\times \Bigg\{ \left[|g_\perp|-\frac{(g^2-g_\perp^2)}{1+r_+}\,C\left(\frac{q^2/(2M)}{-E^\pm_{bs}}\right)\right]^2   \nonumber\\
 && \qquad +\frac{2|g_\perp|(g^2-g_\perp^2)}{1+r_+}\,C\left(\frac{q^2/(2M)}{-E^\pm_{bs}}\right)\nonumber\\
 &&\qquad\times[1-\textrm{sgn}(g_\perp)\sin(2\varphi_{\mathbf q})]\Bigg\}^{1/2}.\label{epmbs}
\eea
The function $C(x)$ appearing in Eq. (\ref{epmbs}) is given by
\bea 
C(  x)&=&-1+\frac{(1+r_+)(1+x)}{r_+x}\ln\left(\frac{1+x}{1+\frac{x}{1+r_+}}\right) , \label{dwav}
\eea
and is such that  $C(x)\geq 0$ $\forall x\geq0$.

 \begin{figure}[b]
\centering
\includegraphics[width=0.95\columnwidth]{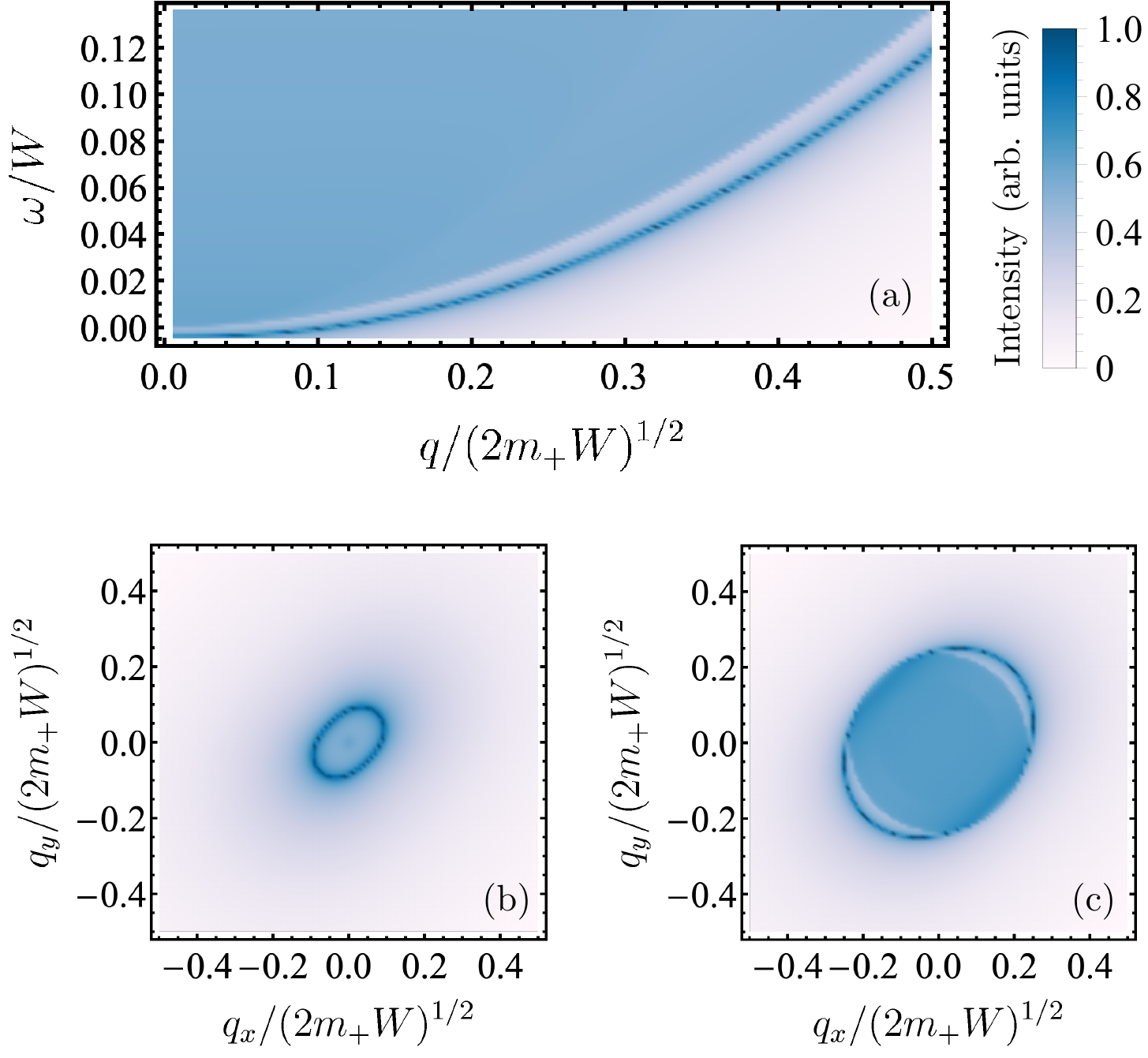}
\caption{Pair spectral function $\mc A_{\textrm{pair}}(\mathbf q,\omega)$ for    $g=0.1$, $g_\perp=-0.5$,  and $r_+=1.2$. In this case,    $|g|<|g_\perp|$ and   only one bound state appears  below the two-particle continuum.  Panel (a) shows $\mc A_{\textrm{pair}}(\mathbf q,\omega)$ as a function of $q$ and $\omega$ at fixed angle   $\varphi_{\mathbf q}=\pi/4$. Panels (b) and (c) show $\mc A_{\textrm{pair}}(\mathbf q,\omega)$ as a function of momentum at fixed $\omega=0$ and $\omega=0.03W$, respectively.  }\label{spectf1}
\end{figure}

For $q\to0$, the result simplifies as $C( x)\sim x\to 0$ and the angle-dependent terms in Eq. (\ref{epmbs}) vanish. In this case, $X_\pm(\mathbf q=0,E_{bs}^\pm)= (1+r_+)/(g\mp |g_\perp|)$ become constant. The  bound state solutions at $\mathbf q=0$, with energies \be
E^\pm_{bs}(\mathbf q=0)=\frac{W}{1-\exp\left(-\frac{1+r_+}{g\mp |g_\perp|}\right)}<0,\label{binding}\ee
exist   as long as $g\pm g_\perp<0$. Therefore, the criterion for the number of bound states at $\mathbf q=0$ matches the three regions depicted in Fig. \ref{rgflow1}. For $g>|g_\perp|$, corresponding to the regime of marginally irrelevant interactions, there are no bound states. We find one bound state with energy $E_{bs}^+$ in  the crossover regime $|g|<|g_\perp|$ and two bound states in the attraction-dominated  regime $g<-|g_\perp|$. For $g<0$ and $g_\perp=0$, the bound states are degenerate at $\mathbf q=0$.  Note also that at weak coupling, $|g|,|g_\perp|\ll1$, the binding energies $E^{\pm}_{bs}(0)\approx -W \exp\left(\frac{1+r_+}{g\mp |g_\perp|}\right)$  are exponentially small, as expected for marginal interactions.   
 
For $g_\perp\neq0$ and $g<-|g_\perp|$,  the bound states     may become degenerate at   nonzero momenta $\mathbf q_0$ such that $X_{+}(\mathbf{q}_0,E_{bs})=X_{-}(\mathbf{q}_0,E_{bs})$. From Eqs. (\ref{Enbs}) and (\ref{epmbs}), we see that the degeneracy point happens along the directions where $\sin(2\varphi_{\mathbf q_0})=\textrm{sgn}(g_\perp)$,
i.e., for angles $\varphi_{\mathbf q_0}=\frac\pi4,\frac{5\pi}4$ for $g_\perp>0$ and $\varphi_{\mathbf q_0}=\frac{3\pi}4,\frac{7\pi}4$ for $g_\perp<0$. The value of $q_0$ is determined by the conditions 
\bea
E_{bs}(  q_0)&=&\frac{\Omega(q_0)}{1-e^{\frac{(1+r_+)g}{g^2-g_\perp^2} }} +\frac{W}{1- e^{-\frac{(1+r_+)g}{g^2-g_\perp^2} }},\\
C\left(\frac{q_0^2/(2M)}{-E_{bs}(q_0)}\right)&=&\frac{(1+r_+)|g_\perp|}{g^2-g_\perp^2}.
\eea

\begin{figure}[t]
\centering
\includegraphics[width=0.95\columnwidth]{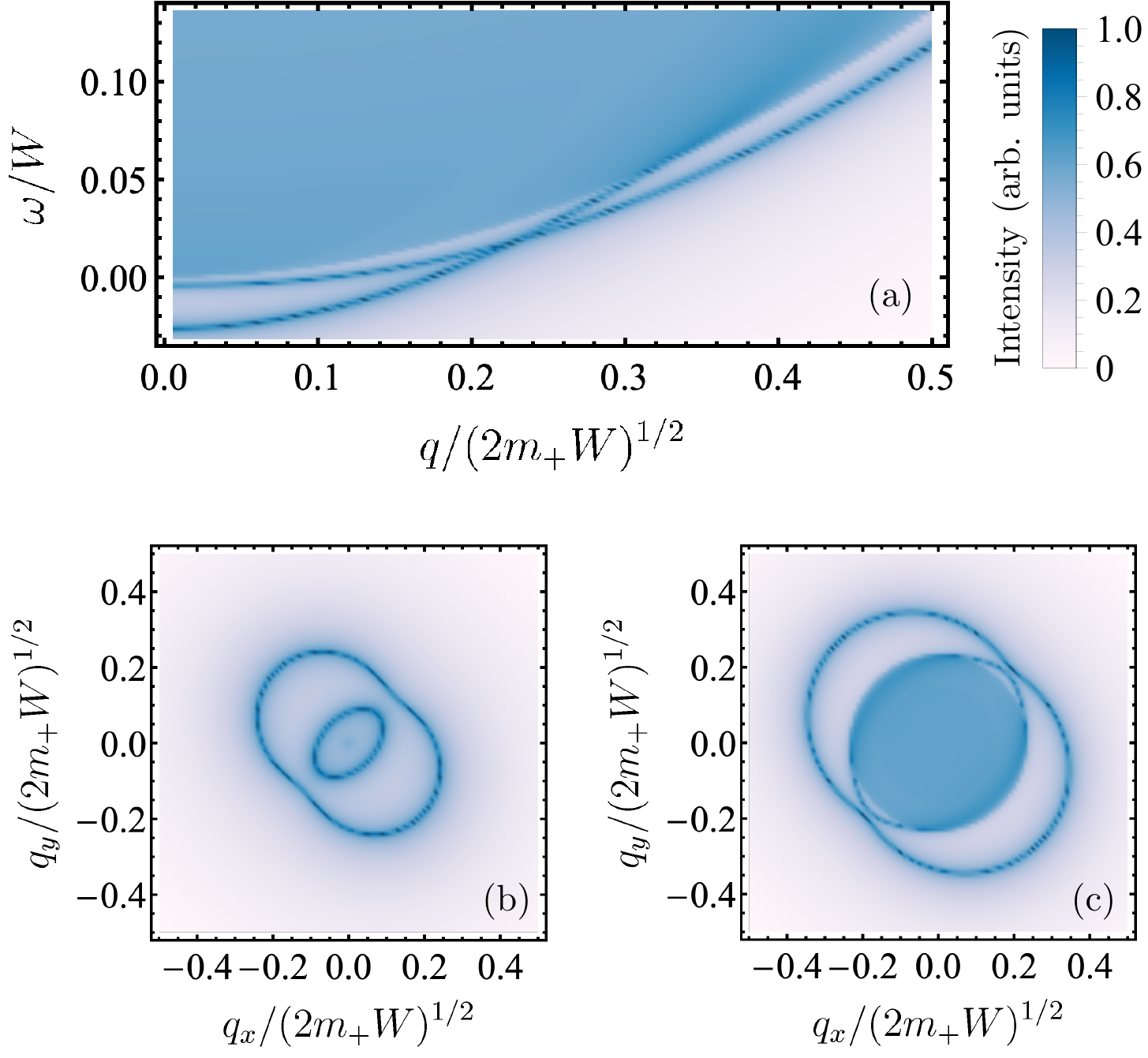}
\caption{Same as Fig. \ref{spectf1} for    $g=-0.5$, $g_\perp=0.1$,  and $r_+=1.2$. In this case,   $g<-|g_\perp|$ and   there are two  bound states   below the continuum  at $\mathbf q=0$.  Note the touching of curves in panel (a), which is due to   the degeneracy of the bound states with momentum $\mathbf q_0\neq 0$ along the direction $\varphi_{\mathbf q}=\pi/4$. }\label{spectf2}
\end{figure}
  
Figures \ref{spectf1} and  \ref{spectf2} show    results for the   pair spectral function in the ladder approximation. The intensities are plotted in logarithmic scale and arbitrary units proportional to  $\ln[1+ A_{\textrm{pair}}(\mathbf q,\omega)/(\eta m_+)]$, with a small    broadening factor $\eta\sim 10^{-4}$. Figure  \ref{spectf1}  is representative of the crossover regime with $|g|<|g_\perp|$. Although the $s$-wave scattering amplitude $g>0$ is repulsive in this example, we do find a bound state below the two-particle continuum. This bound state originates from the effects of anisotropic exchange interaction encoded in $g_\perp$. On the other hand, in the attraction-dominated regime $g<-|g_\perp|$ illustrated by Fig.   \ref{spectf2}, we find two bound states  at $\mathbf q=0$. These bound  states  become degenerate at a finite value of $q$ in the direction $\varphi_{\mathbf q}=\pi/4$, see the anticrossing in Fig. \ref{spectf2}(a). This dependence on $\varphi_{\mathbf q}$ is a manifestation of the unitary transformation in Eq. (\ref{Umatrix}), which is responsible for the nontrivial Berry phase of the QBT point.  Note that the bound state dispersions only exhibit a C$_2$ rotational symmetry,  consistent with the anisotropy of the dipolar exchange interaction in the  lattice model. This contrasts with the   isotropic  single-fermion and impurity dispersions, which account for the  rotational invariance of the edge of two-particle continuum seen in Figs. \ref{spectf1} and  \ref{spectf2}. 

\begin{figure}[t]
\centering
\includegraphics[width=0.95\columnwidth]{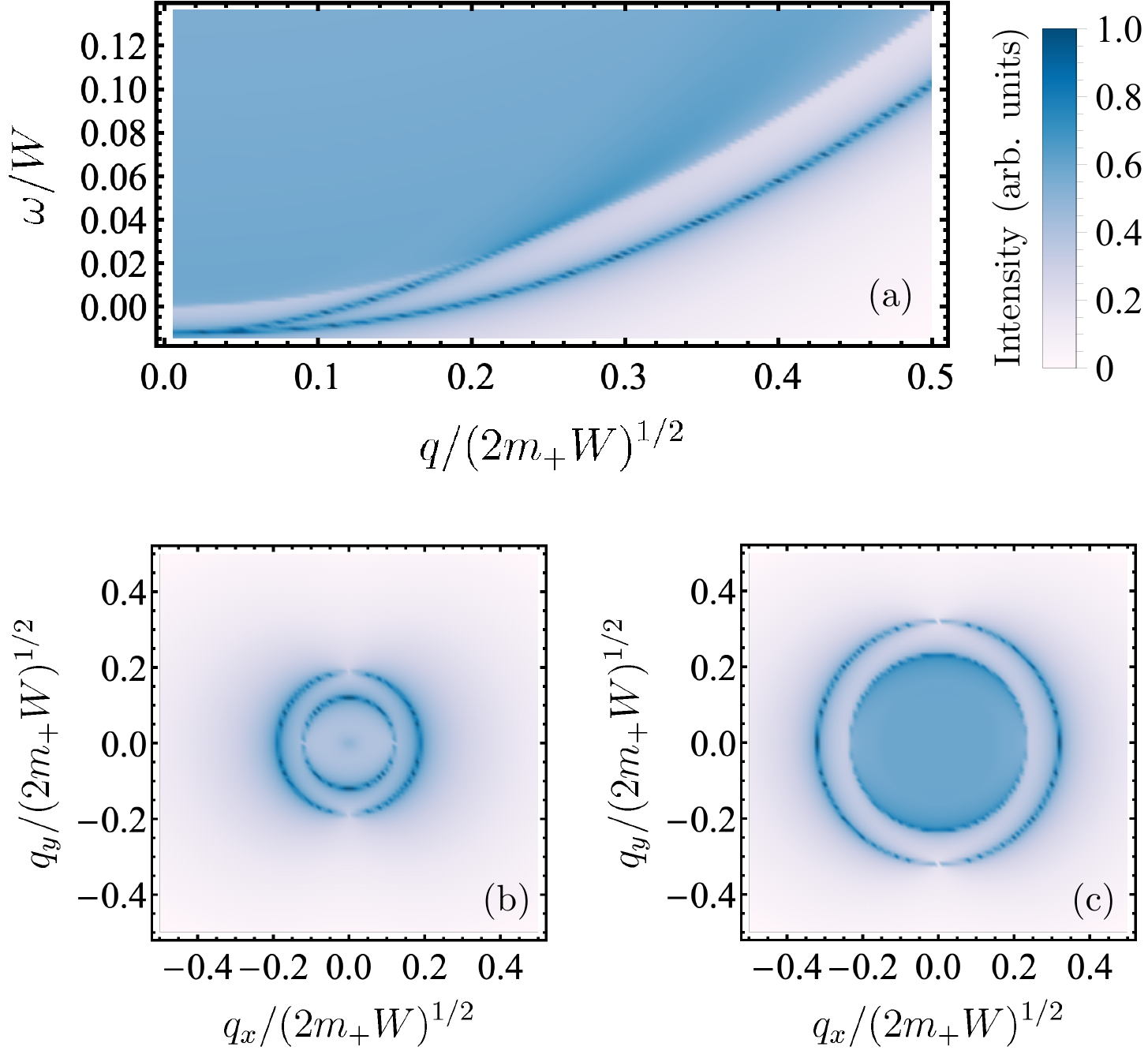}
\caption{A-sublattice component $\mc A^A_{\textrm{pair}}(\mathbf q,\omega)$ of the pair spectral function   for    $g=-0.5$, $g_\perp=0$,  and $r_+=1.2$. Panel (a) shows the result as a function of $q$ and $\omega$ at fixed   $\varphi_{\mathbf q}=\pi/4$. The two bound states become degenerate as $q\to0$.  Panels (b) and (c), taken at fixed $\omega=0$ and $\omega=0.02W$, respectively, show that  the first (second) bound state  has vanishing weight in the A sublattice    for $\varphi_{\mathbf q}=0,\pi$ ($\varphi_{\mathbf q}=\pi/2,3\pi/2$). The angle dependence of the B-sublattice component can be obtained by a C$_4$ rotation of   plots (b) and (c).     }\label{spectf3}
\end{figure}

Finally, consider the case $g<0$ and $g_\perp=0$, which holds  for the standard attractive    Fermi Hubbard model without the dipolar   exchange interaction.  In this case, we are left with the rotationally invariant $g$ interaction. Nevertheless, the bound states can still show signatures of the $d$-wave character of the QBT. Figure \ref{spectf3} displays  the A-sublattice component of the pair spectral function, defined as $ \mc A^A_{\textrm{pair}}(\mathbf q,\omega)=-2\textrm{Im}\{\Pi_{AA}(\mathbf q,\omega+i0^+)\}$. In Fig. \ref{spectf3}(a), we see that the two bound states are degenerate at $\mathbf q=0$, but the degeneracy is lifted as $q$ increases and the second bound state eventually merges with the continuum. Moreover, figures \ref{spectf3}(b) and  \ref{spectf3}(c)  show that the bound state contributions to $ \mc A^A_{\textrm{pair}}(\mathbf q,\omega)$ have nodes as a function of   $\varphi_{\mathbf q}$.  The weight of the first bound state in the A sublattice vanishes for $\varphi_{\mathbf q}=0,\pi$, while for the second bound state it vanishes for $\varphi_{\mathbf q}=\pi/2,3\pi/2$. Along these four directions,  $ \mc A^A_{\textrm{pair}}(\mathbf q,\omega)$ shows only one bound state below the continuum at small $q$.  The location of the nodes is reversed for the  B-sublattice component  of the pair spectral function.  If we add   A and B components, we find that the full pair spectral function is symmetric under C$_4$ rotations, with two bound states in any direction for small $q$. To gain more intuition about the symmetry properties  of the bound states, in Appendix \ref{App2body} we   study the two-body problem of  an impurity interacting with a single particle near the QBT point, without the constraint of a completely filled lower band.

\section{Conclusion\label{sec:conclusion}}

We  studied  the interaction between a    mobile quantum impurity  and a bath of  majority  fermions whose Fermi level is tuned to a quadratic band touching point. The low-energy effective model contains an $s$-wave contact interaction $g$ and a rotational-symmetry-breaking  interaction $g_\perp$ which  can be generated by  dipolar spin  exchange. A renormalization group approach shows a regime in which a  repulsive  impurity-fermion interaction     becomes effectively attractive at low energies. 
This happens because the dipolar spin exchange switches the fermion and the impurity positions, lowering the ground state energy. The amplitude of this process decreases with distance.
This situation leads to the formation of  bound states. The anisotropic momentum dependence of the bound states stems  from the combined  effects of the $g_\perp$ interaction  and the $d$-wave terms in the two-particle propagator.   In the  ladder approximation, we find   a single bound state for $|g|<|g_\perp|$ and two bound states for  $g<-|g_\perp|$, in agreement with the existence of different regimes in the renormalization group flow diagram.  At weak coupling, the binding energies are exponentially small in the coupling constants. 

Higher body bound states, as trimers or tetramers, are not expected to have important contribution to the spectral functions discussed in this work, unless one considers the impurity to be substantially lighter than the fermions and considers the regime of strong interactions, where $p$-wave interactions between the fermions could develop. In addition, the presence of a Fermi sea usually tends to suppress the formation of higher body bound states due to the Pauli exclusion principle, requiring the impurity to be even lighter to allow those bound states \cite{naidon17}.
Our model could be realized with dipolar molecules in an optical checkerboard lattice. 
It should be interesting to extend our results to a low but finite density of minority fermions, with potential implications for unconventional superconductivity in quadratic band touching systems \cite{Pawlak2015}.

\begin{acknowledgments}
We thank T. Enss for helpful discussions. This work is supported by FAPESP/CEPID, FAPEMIG, CNPq, INCT-IQ, and CAPES, in particular through programs CAPES-COFECUB (project 0899/2018) and CAPES-PrInt UFMG (M.C.O.A.). Research at IIP-UFRN is funded by Brazilian ministries MEC and MCTIC. 
\end{acknowledgments}

\appendix

\section{Functions\label{AppFunctions}\label{append}}
In this appendix we write down the expression for the functions $F_i(r_+,r_-)$, with $i=1,\dots,4$, that appear in the RG equations (\ref{RGeqns}). These are given by
\bea
  F_1&=&\int_0^{\pi/2}d\alpha\frac{(1+r_+^{-1})(1+r_-^{-1})}{\sin{\alpha}\cos{\alpha}}\nonumber\\
  & &\times\Bigg[-1+\frac{L}{\left(L^2-\sin^2(2\alpha)\right)^{1/2}}\Bigg], \\
    F_2&=&\frac14\int_0^{\pi/2}d\alpha\,(1+r_+^{-1})(1+r_-^{-1})\sin(2\alpha)\nonumber\\
  & &\times\left[-1+\frac{4L}{\left(L^2-\sin^2(2\alpha)\right)^{3/2}}\right], \\
    F_3&=&2\int_0^{\pi/2}d\alpha\frac{[(1+r_+^{-1})(1+r_-^{-1})]^{3/2}}{(L^2-\sin^2{(2\alpha)})^{3/2}} \nonumber\\
    &&\times\Big\{(1-3L) \sin{(2\alpha)}\nonumber\\
    & &\left.\qquad+\frac{2\big[(L^2-\sin^2{(2\alpha)})^{3/2}-L^3\big]}{\sin{(2\alpha)}}\right\}, \\
    F_4&=&\int_0^{\pi/2}d\alpha\frac{[(1+r_+^{-1})(1+r_-^{-1})]^{3/2}}{(L^2-\sin^2{(2\alpha)})^{5/2}}\sin{(2\alpha)} \nonumber\\
    & &\times\big[(1-3L)\sin^2{(2\alpha)}+2L^2\big], 
\eea
where 
 \begin{eqnarray}
    L(\alpha)&=&(1+r_-^{-1})\cos^2{\alpha}+(1+r_+^{-1})\sin^2{\alpha}. 
 \end{eqnarray}

\section{Two-body problem \label{App2body}}
In this appendix we consider the two-body problem described by the Schr\"odinger equation\bea
E\Phi(\mathbf r_1,\mathbf r_2)&=&\left[h_0(\mathbf r_1)-\frac{\openone}{2M}\nabla_{\mathbf r_2}^2\right]\Phi(\mathbf r_1,\mathbf r_2)\nonumber\\
&&+\delta(\mathbf r_1-\mathbf r_2)(g\openone+g_\perp\sigma^x)\Phi(\mathbf r_1,\mathbf r_2),\label{schrod}
\eea
where $\Phi(\mathbf r_1,\mathbf r_2)$ is the wave function with the first particle representing the fermion near the QBT and the second particle representing the impurity. In addition to the dependence on the coordinates $\mathbf r_1$ and $\mathbf r_2$, the wave function contains a spinor in sublattice space for the first particle. Taking the Fourier transform of Eq. (\ref{schrod}), we obtain\bea
E\tilde\Phi(\mathbf p_1,\mathbf p_2)&=&\left[h_0(\mathbf p_1)+\openone\frac{p^2_2}{2M}\right]\tilde\Phi(\mathbf p_1,\mathbf p_2)\nonumber\\
&&+\int\frac{d^2q}{(2\pi)^2}(g\openone +g_\perp\sigma^x) \tilde \Phi(\mathbf p_1+\mathbf q,\mathbf p_2-\mathbf q).\nonumber\\
\eea

Let us focus on the case $\mathbf P=\mathbf p_1+\mathbf p_2=0$, corresponding to vanishing center-of-mass momentum. We then define\be
\Delta(\mathbf p)=\left[\left(E-\frac{p^2}{2M}\right)\openone-h_0(\mathbf p)\right]\tilde\Phi(\mathbf p,-\mathbf p),\label{bigDelta}
\ee
and obtain\bea
\Delta(\mathbf p)&=&\int\frac{d^2q}{(2\pi)^2}(g\openone +g_\perp\sigma^x)\nonumber\\
&&\times\left[\left(E-\frac{q^2}{2M}\right)\openone-h_0(\mathbf q)\right]^{-1}\Delta(\mathbf q).\label{gapeqn}
\eea
Since the right-hand side of Eq. (\ref{gapeqn}) does not depend on $\mathbf p$, we have that $\Delta(\mathbf p)=\Delta_0$ is  a constant spinor. Thus, Eq. (\ref{gapeqn}) reduces to the eigenvalue equation\be
\mc R\Delta_0=\Delta_0,\label{RDelta}
\ee
where\bea
 \mc R&=&\int\frac{d^2q}{(2\pi)^2}(g\openone +g_\perp\sigma^x)\nonumber\\
&&\times\left[\left(E-\frac{q^2}{2M}\right)\openone-h_0(\mathbf q)\right]^{-1}.\label{eigenvalue}
\eea

To solve Eq. (\ref{eigenvalue}),  we use the unitary transformation that diagonalizes $h_0(\mathbf q)$ and perform the integral in the disc  $0<q<(2m_+W)^{1/2}$ with high-energy cutoff $W$.   We find that bound state solutions with $E=\mc E_{bs}<0$ exist only if $m_->M$. This condition is not satisfied for the lattice model discussed in Sec. \ref{model}, but more generally one could modify the band structure by adding further hopping processes or make the impurity out of another  atomic species  with a different  mass. At weak coupling, the binding energies scale as \be
\mc E^{\pm}_{bs}\sim - W\exp\left[\frac{m_+}{(\bar \mu_++\bar\mu_-)(g\pm g_\perp)}\right],
\ee
where  $\bar \mu_+=\mu_+$ and $\bar \mu_-=m_-M/(m_--M)$. The bound states are degenerate for $g_\perp=0$. If $g_\perp\neq0$, there is no bound state for $g>|g_\perp|$, one bound state for $|g|<|g_\perp|$ and two bound states for $g<-|g_\perp|$. This result is equivalent  to the criterion for bound states in the many-body problem. 

We obtain the bound state wave functions for $\mathbf P=0$ by substituting the   eigenvectors $\Delta_0$ from Eq. (\ref{RDelta}) into Eq. (\ref{bigDelta}). In the regime where the bound states exist, we have \bea
\tilde\Phi_\pm(\mathbf p,-\mathbf p)&= &\mc N\Big\{f_s(p,\mc E_{bs}^\pm)\openone+f_d(p,\mc E_{bs}^\pm )\left[\cos(2\varphi_{\mathbf p})\sigma^z\right. \nonumber\\
&&\left. +\sin(2\varphi_{\mathbf p})\sigma^x\right]\Big\} \left(\begin{array}{c}1\\\pm 1\end{array}\right),
\eea
where $\mc N$ is a normalization factor. The functions \bea
f_s(p,\mc E)&=&\left(\frac{p^2}{2\bar \mu_+}-\mc E\right)^{-1}+\left(\frac{p^2}{2\bar \mu_-}-\mc E\right)^{-1},\nonumber\\
f_d(p,\mc E)&=&\left(\frac{p^2}{2\bar \mu_+}-\mc E\right)^{-1}-\left(\frac{p^2}{2\bar\mu_-}-\mc E\right)^{-1}, \label{sdcomp}
\eea
represent the amplitudes  of the $s$- and $d$-wave components of the bound state wave function, respectively. Note that $f_d(p,\mc E)$ vanishes for $p\to0$. At nonzero $p$, we can write $\tilde\Phi_\pm(\mathbf p,-\mathbf p)=\chi_\pm(p,\varphi_{\mathbf p})$, with the symmetry properties \bea
\chi_\pm\left(p,\varphi_{\mathbf p}+\frac\pi4\right)&=&\pm \sigma^x \chi_\pm\left(p,-\varphi_{\mathbf p}+\frac\pi4\right),\nonumber\\
\chi_\pm\left(p,\varphi_{\mathbf p}-\frac\pi4\right)&=&\pm \sigma^x \chi_\pm\left(p,-\varphi_{\mathbf p}-\frac\pi4\right). 
\eea
For $g_\perp=0$, the bound states become degenerate, $\mc E_{bs}^+=\mc E_{bs}^-$, and we have 
\be
i\sigma^y\chi_\pm\left(p,\varphi_{\mathbf p}+\frac\pi2\right)=\pm \chi_\mp(p,\varphi_{\mathbf p}).
\ee
In this case we can take linear combinations of   $ \chi_+(p,\varphi_{\mathbf p})$ and $ \chi_-(p,\varphi_{\mathbf p})$ to form  eigenstates of the C$_4$ rotation.  

Both $s$- and $d$-wave components in Eq. (\ref{sdcomp}) have a Lorentzian dependence on $p$. This implies an exponential decay as a function of the relative distance $r=|\mathbf r_1-\mathbf r_2|$ in real space, with length scales $ \sim (\bar\mu_\pm |\mc E_{bs}|)^{-1/2}$.

%
%\begin{figure}[h]
%\centering
%\includegraphics[width=0.85\columnwidth]{Figuras/spectfunc_m22_M10_U-1_J-1,5_n10-3_W1_qx0,1.png}
%\caption{Spectral function using $m_+=2, M=10,g_U=-1,g_J=-1.5,\eta=10^{-3},W=1,\bm{q}=(0.1,0)$.}\label{specfunc1}
%\end{figure}
%\begin{figure}[h]
%\centering
%\includegraphics[width=0.85\columnwidth]{Figuras/spectfunc_m22_M10_U-4_J-1_n10-2_W1_qx0,1.png}
%\caption{Spectral function using $m_+=2, M=10,g_U=-4,g_J=-1,\eta=10^{-2},W=1,\bm{q}=(0.1,0)$.}\label{specfunc2}
%\end{figure}
%

%\nocite{*}

\bibliographystyle{apsrev4-1}
\bibliography{references}% Produces the bibliography via BibTeX.

\end{document}